# Texture features in medical image analysis: a survey

**Faeze Kiani**[1,*]

[1]Department of electronic science, Online Computer Vision Research Group, Iran
* Corresponding author:  ocvrgroup000@gmail.com

**Abstract:** The texture is defined as spatial structure of the intensities of the pixels in an image that is repeated periodically in the whole image or regions, and makes the concept of the image. Texture, color and shape are three main components which are used by human visual system to recognize image contents. In this paper, first of all, efficient and updated texture analysis operators are survived with details. Next, some state-of-the-art methods are survived that use texture analysis in medical applications and disease diagnosis. Finally, different approaches are compared in terms of accuracy, dataset, application, etc. Results demonstrate that texture features separately or in joint of different feature sets such as deep, color or shape features provide high accuracy in medical image classification.

## 1. Introduction

Today, with the growth of information technology and hardware facilities, artificial intelligence is widely used in various scientific fields such as medicine, agriculture, construction, tourism, biochemistry, etc. Among the different branches of artificial intelligence science, image processing and computer vision include a large part of the applications [1]. The ultimate goal in all computer vision-based systems is to perform activities that the human eye is capable of performing more quickly and accurately. Recent research shows that the human eye usually uses three components of color, texture and shape to recognize the nature of images. Meanwhile, texture plays a more important role [2]. Therefore, so far, several operators have been presented for analyzing the texture of images and extracting texture features in the image.

The texture of an image means the spatial structure of the brightness intensity of the pixels, which is periodically repeated in the whole image or parts of it, and makes the nature of the image [3]. The concept of image texture refers to the natural concept of the object in the image. In fact, objects in daily life have texture. Even humans try to repeat a texture to make things. For example, a brick wall has a repeating texture that is made by humans. Also, the body structure of a tiger has a repeating pattern and texture. Even the monitor of a mobile device or the way the keyboard is displayed has a certain pattern that creates the texture of that image and makes it possible for humans to recognize that object.

Identification and analysis of image texture is used in various fields, including medicine [4], agriculture [5], industry [6], tourism [7], biochemistry [8], etc. Usually, in many diseases, the desired organ in the body undergoes changes in appearance after infection. These changes can be clearly seen in the texture of the image. Therefore, image texture analysis techniques are used for medical diagnosis and medical image analysis. In this article, some successful methods in the analysis of medical images based on texture analysis are reviewed. In the following, first in the next section, efficient operators for image texture analysis in the field of image processing are reviewed. Then, in the third part, the common uses of tissue in medicine are reviewed based on published articles. An example of natural texture and intelligent textural images are shown in the figure 1.

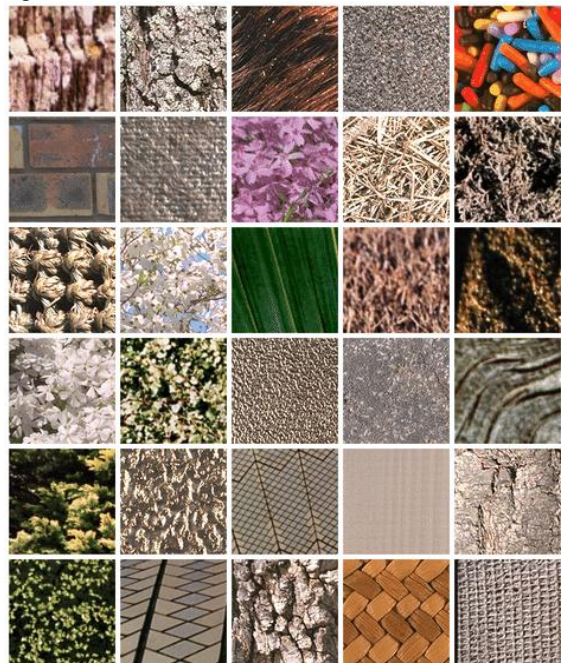

**Figure 1.** Natural and intelligent texture images

## 2. Texture analysis techniques

Until now, various methods have been presented for image texture analysis. Most of the methods presented in this field try to discover the repeating pattern in the image and extract it in the form of numerical features so that machine learning algorithms can be used for object identification [9], defect detection [10], quality control [11], medical diagnosis [12], etc.

One of the simplest methods of extracting image texture features is image histogram analysis [13]. Image histogram is a two-dimensional graph where the x-axis shows the gray level of the image and the vertical axis shows the number of pixels with the desired gray level in the image. Therefore, each bin (column) in the normalized histogram shows the probability of occurrence of the desired gray level in the input image. The normalized histogram of the image is a statistical and probabilistic graph, so different features can be extracted from it, each of which defines a specific feature (properties) of the graph. Some of the popular texture features which can be extracting from histogram are as follows:

Mean: $\quad f_1 = \sum_{i=0}^{G-1} iP(i)$ (1)

Skewness: $\quad f_2 = \sigma^{-3} \sum_{i=0}^{G-1} (i - f_1)^3 p(i)$ (2)

Kurtosis= $\quad f_3 = \sigma^{-4} \sum_{i=0}^{G-1} (i - f_1)^4 p(i) - 3$ (3)

Energy = $\quad f_4 = \sum_{i=0}^{G-1} [p(i)]^2$ (4)

Entropy= $\quad f_5 = \sum_{i=0}^{G-1} p(i) \log_2[p(i)]$ (5)

Where, P(i) shows the occurence probability of gray-level i in the image (the height of $i^{th}$ bin in the normalized histogram). Also, G is the total number of possible gray-levels in the image which can be considered as 256 in images with 8 bits format.

Gray-level Co-occurrence matrices (GLCM) are one of the most efficient image texture analysis operators [14]. Each cell in the co-occurrence matrix with (x, y) coordinates represents the number of times in the image that the gray level "x" has been observed under the specific relationship (d) with the gray level "y" in the image. By dividing all the cells of the GLCM by the total number of pixels, the normalized co-occurrence matrix is obtained. An example of GLCM process is shown in the figure 2.

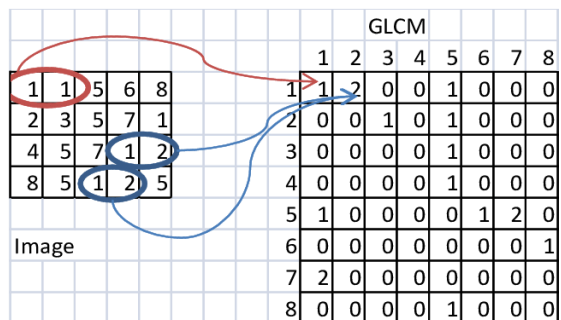

**Figure 2**. An example of GLCM process

The GLCM is also a statistical and probabilistic operator. Therefore, extracting statistical features from it can represent the texture of the image.

Some of the texture features which can be extracted from the GLCM are as follows.

| Feature | Description | |
|---|---|---|
| Energy (ENR) | $ENR = \sum_{i=0}^{n-1} \sum_{j=0}^{n-1} G(i,j)^2$ | (6) |
| Entropy (ENT) | $ENT = -\sum_{i=0}^{n-1} \sum_{j=0}^{n-1} G(i,j) \ln[G(i,j)]$ | (7) |
| Contrast (CON) | $CON = \sum_{i=0}^{n-1} \sum_{j=0}^{n-1} G(i,j)(i-j)^2$ | (8) |
| Difference entropy (DENT) | $DENT = -\sum_{i=0}^{n-1} G_{x-y}(i) \ln[G_{x-y}(i)]$ | (9) |
| Difference variance (DVAR) | $DVAR = -\sum_{i=0}^{n-1} G_{x-y}(i)(i-DENT)^2$ | (10) |
| Maximum probability (MAXP) | $MAXP = MAX_{i,j} G(i,j)$ | (11) |
| Sum entropy (SENT) | $SENT = -\sum_{i=2}^{2n} G_{x+y}(i) \ln[G_{x+y}(i)]$ | (12) |
| Sum average (SVAR) | $SVAR = \sum_{i=0}^{n-1} iG_{x+y}(i)$ | (13) |
| Homogeneity (HOM) | $HOM = \sum_{i=0}^{n-1} \sum_{j=0}^{n-1} G(i,j)/(1+(i-j)^2$ | (14) |
| Correlation (COR) | $COR = \sum_{i=0}^{n-1} \sum_{j=0}^{n-1} \frac{G(i,j)[(i-\mu_x)(j-\mu_y)]}{\sigma_x \sigma_y}$ | (15) |

Local binary pattern (LBP) is one of the most widely used operators in the field of image texture analysis [15]. This operator generates a binary pattern for each pixel of the image based on the relationship between the intensity of that pixel and its neighbors. To execute this operator, first a neighborhood is considered for the pixel in question, and the center of the neighborhood will be the desired pixel. Then, the intensity of each one of the neighbors is compared with the center of the neighborhood, and if the intensity of the neighbor is greater than the center, that point is indicated by bit 1, otherwise it is indicated by bit zero. Finally, a binary pattern is produced around each pixel. If the number of neighbors is eight, the generated binary pattern will have 8 bits and it can be converted to a number in base of ten. So finally, if the number of neighbors is 8, the numeric value of the LBP for each pixel is between 0 and 255. After applying the local binary pattern operator on the whole image, a histogram of LBP values can be produced and used for feature extraction. An example of LBP process is shown in the figure 3.

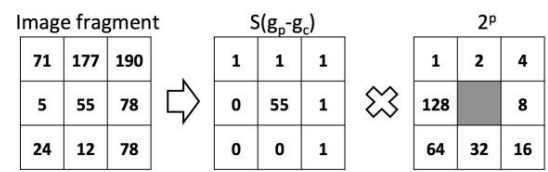

LBP= 1 x 1 + 1 x 2 + 1 x 4 + 1 x 8 + 1 x 16 + 0 x 32 + 0 x 64 + 0 x 128 = 4 + 8 + 16 = 31
**Figure 3**. An example of LBP process

The way of choosing neighbors can be different, for example in [16], the authors proposed for the first time one-dimensional local binary patterns that are used in the field of visual inspection systems. Ojala et al. [17], presented a new algorithm for feature extraction from locally extracted binary patterns, which is known as the Modified local binary patterns (MLBP). For the first time, Tan et al. [18], defined the difference in intensities of pixels based on three component (-1, 1, and 0) and in this regard, they presented the operator of local ternary patterns (LTP). An example of LTP process is shown in the figure 4.

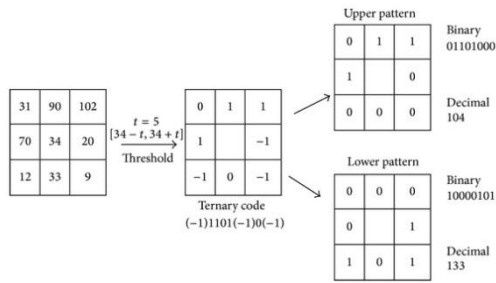

**Figure 4**. An example of LTP process

The operator of five local patterns was proposed by Nanni et al. [19] to reduce sensitivity to noise and increase the resolution of extracted features. Armi et al. [20], presented an improved version of local quintuple patterns that dynamically calculates the thresholds of generating patterns. It is known as improved local quinary patterns (ILQP). Local neighborhood difference patterns (LNDP) is an innovative lbp-like operators which consider the relation between neighbors in joint of their relation with center pixel [21]. Multi Threshold Uniform-based Local Ternary Patterns (MT_ULTP) is proposed for cell phenotype classification, which remove non-uniform local patterns before feature extraction [22].

The gray level run length matrix (GLRLM) is also a statistical operator that is used for image texture analysis. In this matrix, the rows represent the gray levels and the columns show the length of the gray level. For example, the cell value with coordinates (2, 4) in this matrix shows that how times the gray level 2 has been observed with a fixed length of 4 pixels in the image. An example of GLRM process is shown in the figure 5.

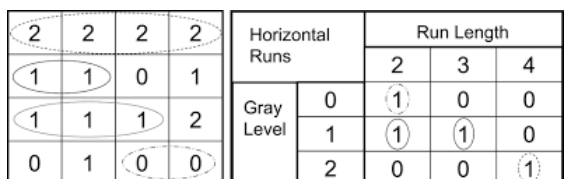

**Figure 5**. An example of GLRLM process

Some researchers have tried to transfer the image from the spatial environment to the serious environment by using transformation functions and provide operators to extract the texture features in the new environment. For example, wavelet functions transfer the image to the frequency domain. Then, in this environment, wavelet coefficients are extracted as features. Some methods extract texture features by using image filtering, among which Gabor filters or SIFT filters can be mentioned.

## 3. Texture features in medical analysis

Zeebaree et al. [23] performed uniform LBP in different radius sizes for breast cancer diagnosis. The general framework of the proposed approach in [23], is shown in the figure 6. As mentioned in the figure 6, the LBP is performed with different threshold such as center intensity, mean of neighbors, upper possible intensity between neighbors and lower value to build four different LBP histograms to extract texture features.

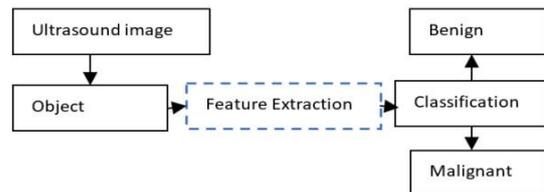

**Figure 6.** The general framework of the breast cancer diagnosis system in [23]

Bruno et al. [24] used combination of LBP and Curvelet transform for breast cancer classification. The main block diagram of the proposed approach in [24] is shown in the figure 7. First of all ultrasound breast image is transformed to curvelet domain. Next, LBP is performed on transformed images in different scales. Also, to reduce complexity, finally feature selection technique is used

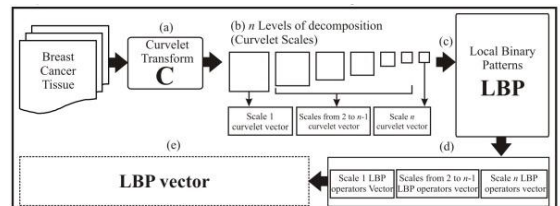

**Figure 7.** The main block diagram of breast cancer diagnosis system in [24]

The performance of LBP and local gradient patterns (LGP) as two efficient texture analysis operators is presented in [25] for breast cancer classification. Texture features play important role in cytoplasm analysis in cervix cells. Modified uniform local ternary patterns (MULTP) as feature extractor and multi-layer perceptron (MLP) as classifiers are used in for cervical cancer diagnosis [26].

Arya et al. [27], employed a set of texture feature for pap smear image classification. The proposed approach consists of two stages, feature extraction and classification. First of all, popular handcrafted texture features such as gray-level co-occurrence matrixes (GLCM), local binary patterns (LBP), Law's and discrete wavelet transform (DWT) are extracted from input single cell image. Next, linear support vector machine (SVM) is used for classification stage.

Texture statistical features extracted from histogram in joint of global significant value (GSV) and time-series information are performed in a concatenating format for pap smear classification [28]. Pap smear image is an efficient tool for human experts to detect cervical cancer.

DNA image cytometry is widely used in cytopathology as a means to obtain objective information concerning the diagnosis and prognosis of human cancer. A combination of texture features including descriptive statistical features, chromatin distribution, range extreme, markov model, GLRLM and fractal texture features are used in [29] for DNA image cytometry. Texture image analysis can be used in different DNA issues such as DNA repair gene process [30].

Lung cancer tumor detection and segmentation is another branch which texture feature can be used widely. In [31], the basic GLCM operator is used for lung cancer classification. GLCM is performed in different directions to improve final classification accuracy. In [33] texture feature are used in combination of active contours and super pixel techniques to detect lung cancer in CT-scan image of chest. Antifungal Activity detection is another branch which texture analysis can be useful for automatic activity detection [34]. Different studies can be used about skin diseases detection which uses texture analysis. In [35], skin cancer is performed based on texture features extracted from natural skin image. Lupus erythematosus [36] is one of the skin diseases whose early diagnosis can help to increase the efficiency of treatment. Texture features can be very efficient in analyzing the skin of a patient. Some studies can be found in cell and bacteriophages image analysis which use texture operators. For example, Nanni et al. [37] compare the performance of popular variants of LBP texture analysis operators for cell phenotype classification. Basic LBP, LTP, LPQ and ILBP are compared in [37]. As another example, multi-threshold uniform local ternary pattern (MT-ULTP) is presented in [38] to classify cell types. The analysis of the isolated bacteriophages from sewage water [39] is another branch which can be done using texture analysis. Texture features play role in prediction and analysis of H. pylori infection prevalence. Different computer vision-based approaches have been proposed in this scope since now [40-43].

Amin et al. [44] performed fusion of different texture analysis operators such as LBP, histogram of gradient (HOG) and SFTA for brain tumor detection.

Biomedical image processing using Magnetic Resonance Imaging (MRI) is efficient tool for brain tumor detection. The main aim of [45] is to classify the brain into two classes, tumor or healthy. Contourlet Transformation is used for Region of Interest (ROI) segmentation. Next, feature extraction is applied using GLRLM and Center-Symmetric Local Binary Patterns (CSLBP).

## 4. Results

This is a review article regarding the application of image texture analysis in medicine. Therefore, in this section, based on the results reported in the survived papers, the effect of image texture analysis in various medical applications is examined. In relation to the survived papers, various applications such as breast cancer, skin cancer and lung tumor have been considered. The results of the review of articles in Table 1 are compared based on the type of problem, the type of texture operator, dataset and classification accuracy.

**Table 1.** Overview of texture operators in medical applications

| Ref. | Problem | Operator | Dataset | Acc.(%) |
|---|---|---|---|---|
| [23] | Breast cancer | ULBP | Self-collected | 96.1 |
| [24] | Breast cancer | LBP + Curvelet | DDSM | 94 |
| [24] | Breast cancer | LBP + Curvelet | UCSB | 100 |
| [26] | Cervical cancer | ULTP | Herlev | 98.69 |
| [27] | Cervical cancer | GLCM | Herlev | 83.4 |
| [27] | Cervical cancer | LBP | Herlev | 84.8 |
| [27] | Cervical cancer | DWT | Herlev | 77.6 |
| [27] | Cervical cancer | GLCM+LBP+DWT | Herlev | 88.2 |
| [28] | Cervical cancer | SF+GSV+IQR | Herlev | 88.45 |
| [31] | Lung tumor | GLCM | Self-collected | 81.8 |
| [32] | Lung tumor | GLCM + Fractal | FDG | 96.1 |
| [35] | Skin cancer | Color | DermIS | 83.5 |
| [35] | Skin cancer | LBP | DermIS | 87.35 |
| [35] | Skin cancer | Color + LBP | DermIS | 90.32 |
| [37] | Cell classification | EQP | 2d-Hela | 92.4 |
| [37] | Cell classification | LBP | 2d-Hela | 82.3 |
| [37] | Cell classification | LTP | 2d-Hela | 91.4 |
| [38] | Cell classification | MT-ULTP | 2d-Hela | 86.77 |

## 5. Conclusion

The main purpose of this article was to investigate image texture analysis operators in the field of medicine. In this regard, first, some of the most efficient operators and techniques for extracting image texture features were reviewed. In the following, their application in different branches of medical science was investigated. The results reported in showed that image texture operators can interpret all types of medical images such as MRI, CT scan, PET scan well. Also, the results showed that different supervised classifiers can exactly joined by texture features for medical image classification. This problem is used to solve some challenges in the medical field, such as tumor diagnosis. Today, researchers are trying to use features such as color, deep, geomtric, and shape in addition to texture features to increase the final accuracy of the system. But ignoring image texture information in medicine greatly reduces the final efficiency of the system.